\documentstyle[12pt]{article}
\def\ie{{\em i.e.}}

\catcode`\@=11 % This allows us to modify PLAIN macros.
\def\coeff#1#2{{\textstyle{#1\over #2}}}

\def\vev#1{\langle #1\rangle}
\def\lsim{\mathrel{\mathpalette\@versim<}}
\def\gsim{\mathrel{\mathpalette\@versim>}}
\def\@versim#1#2{\vcenter{\offinterlineskip
    \ialign{$\m@th#1\hfil##\hfil$\crcr#2\crcr\sim\crcr } }}
\def\etal{{\em et. al.}}
\def\JL{J. L. Lopez}
\def\DVN{D. V. Nanopoulos}

\def\r#1{$\bf#1$}
\def\rb#1{$\bf\overline{#1}$}
\def\GeV{\,{\rm GeV}}

\def\NPB#1#2#3{Nucl. Phys. B {\bf#1} (19#2) #3}
\def\PLB#1#2#3{Phys. Lett. B {\bf#1} (19#2) #3}

\def\PRD#1#2#3{Phys. Rev. D {\bf#1} (19#2) #3}
\def\PRL#1#2#3{Phys. Rev. Lett. {\bf#1} (19#2) #3}

\def\hepph#1{{\tt hep-ph/#1}}
\def\hepth#1{{\tt hep-th/#1}}

\begin{document}
\begin{flushright}
\baselineskip=12pt
DOE/ER/40717--20\\
CTP-TAMU-45/95\\
ACT-16/95\\
\tt hep-ph/9511426
\end{flushright}

\begin{center}
\vglue 1cm
{\Large\bf A New Scenario for String Unification\\}
\vglue 0.75cm
{\Large Jorge L. Lopez$^1$ and D.V. Nanopoulos$^{2,3}$\\}
\vglue 0.5cm
\begin{flushleft}
$^1$Department of Physics, Bonner Nuclear Lab, Rice University\\ 6100 Main
Street, Houston, TX 77005, USA\\
$^2$Center for Theoretical Physics, Department of Physics, Texas A\&M
University\\ College Station, TX 77843--4242, USA\\
$^3$Astroparticle Physics Group, Houston Advanced Research Center (HARC)\\
The Mitchell Campus, The Woodlands, TX 77381, USA
\end{flushleft}
\end{center}

\vglue 0.75cm
\begin{abstract}
We present a new scenario for gauge coupling unification in flipped SU(5)
string models, which identifies the $M_{32}$ scale of SU(3) and
SU(2) unification with the empirical $M_{\rm LEP}\sim10^{15-16}$~GeV scale,
and the $M_{51}$ scale of SU(5) and U(1) unification with the theoretical
$M_{\rm string}\sim5\times10^{17}$~GeV string unification scale. The vacuum
shift necessary for the cancellation of the anomalous $\rm U_A(1)$ and an
SU(4) hidden sector with fractionally-charged particles, play a crucial role in
the dynamical determination of all intermediate mass scales in this scenario.
\end{abstract}
\vspace{1cm}
\begin{flushleft}
\baselineskip=12pt
DOE/ER/40717--20\\
CTP-TAMU-45/95\\
ACT-16/95\\
November 1995
\end{flushleft}
\newpage

\setcounter{page}{1}
\pagestyle{plain}
\baselineskip=14pt

The convergence of the Standard Model gauge couplings, when extrapolated to
very high energies in the context of supersymmetric theories, has received a
great deal of attention ever since it was first observed \cite{amaldi-costa},
and especially since the advent of the LEP era \cite{EKN}. In the simplest and
best studied scenarios, only the particles in the Standard Model and their
superpartners are included in the evolution of the gauge couplings, which are
seen to converge at the scale
\begin{equation}
M_{\rm LEP}\sim10^{15-16}\GeV\ ,
\label{eq:MLEP}
\end{equation}
above which a larger structure must be revealed. As compelling as this simple
result may be, it is more than one order-of-magnitude lower than the scale at
which the gauge couplings should unify \cite{Ginsparg} (to lowest order) in the
context of superstring models \cite{Kaplunovsky},
\begin{equation}
M_{\rm string}=5\times g\times10^{17}\GeV\ ,
\label{eq:Mstring}
\end{equation}
where $g$ is the unified gauge coupling at this scale. This discrepancy of
scales (which may enhance the reliability of the low-energy effective theory
description) has been taken seriously by string model builders, as $M_{\rm
LEP}$ is an empirical result, whereas $M_{\rm string}$ is an actual theoretical
prediction. Early on it was pointed out that reconciliation of these two
scales required to supplement the particle content of the Minimal
Supersymmetric Standard Model (MSSM) with new intermediate-scale particles
\cite{price}. Alternatively, it was proposed that threshold corrections from
the infinite tower of massive string states shifted $M_{\rm string}$
down to effectively coincide with $M_{\rm LEP}$ \cite{Ibanez}. The latter
scenario is now disfavored, as it requires (large) values of the (moduli)
fields that parametrize the threshold corrections, which are hard to obtain in
actual string models \cite{moduli,DF}. Moreover, this scenario appears to
push ``gravity" down to scales uncomfortably lower than the Planck mass, and
still requires the addition of new particles beyond the MSSM \cite{DF}. Further
generic alternatives may exist \cite{Other}, but these are yet to be realized
in realistic string models.

We then see that all known scenarios of string unification predict the
existence of new intermediate-scale particles in the observable sector, a
property quite common among actual string models, such as those based on
the gauge groups SU(5)$\times$U(1) (``flipped" SU(5)), or
SU(3)$\times$SU(2)$\times$U(1) (``standard-like" models), or
SU(4)$\times$SU(2)$\times$SU(2) (``Pati-Salam" models), or $\rm SU(3)^3$.
Of these possible gauge groups, only flipped SU(5) unifies the SU(3) and SU(2)
non-abelian factors of the Standard Model gauge group (at the scale $M_{32}$),
and therefore can in principle reproduce the ``observed" $M_{\rm LEP}$ scale
(identified with $M_{32}$), above which SU(5)$\times$U(1) is revealed. Note
that this result is unaffected (to lowest order) by the introduction of the
intermediate-scale representations, if these come in complete SU(5) multiplets
as we advocate below. The SU(5) and U(1) gauge couplings unify at the scale
$M_{51}$, which we consistently identify with $M_{\rm string}$. In the case of
the other gauge groups, unification must occur at $M_{\rm string}$ leaving no
room for $M_{\rm LEP}$, which must be regarded as an accidental result.

In this Letter we present a new scenario for string unification in the context
of flipped SU(5) models, following the guidelines just described.
Such models have been derived from string \cite{revamp,search} and, through
detailed first-principles calculations
\cite{decisive,LN,Erice95}, have been shown to possess many interesting
properties, which even though not of crucial importance for the subsequent
discussion, do motivate further the consideration of this class of models.
The latest incarnation of the string model \cite{search} includes three
generations of quarks and leptons, an SU(5)$\times$U(1) observable gauge group,
an SO(10)$\times$SU(4) hidden sector gauge group, vanishing vacuum energy at
tree-level ($V_0=0$), and vanishing quadratically-divergent one-loop correction
to the vacuum energy ($Q=0$) in the shifted vacuum where the anomalous $\rm
U_A(1)$ is cancelled to ensure unbroken supersymmetry at the string scale.
Moreover, in this vacuum shifting the SU(5)$\times$U(1) symmetry is broken, and
one is able to find naturally solutions \cite{Erice95} with $M_{32}\sim M_{\rm
LEP}$, as advocated above.

Crucial to the string unification program are a pair of (\r{10},\rb{10}) SU(5)
representations, in addition to those required for SU(5)$\times$U(1) symmetry
breaking, with intermediate-scale masses $M_{10}$. In fact, in the class of
fermionic string models that we study, the vanishing of the tree-level vacuum
energy ($V_0=0$) appears inextricably correlated to the existence of the extra
(\r{10},\rb{10}) pair \cite{LN}. Also important are the (\r{4},\rb{4})
representations of the hidden SU(4) gauge group, which have fractional electric
charges ($\pm{1\over2}$), and are either heavy or become confined into
integrally-charged ``cryptons" at the SU(4) confinement scale $\Lambda_4$
\cite{cryptons}. Moreover, a condensate of such hidden sector fields provides
the mass scale that determines $M_{10}$ dynamically. The novelty in our
approach is that all intermediate scales, including $M_{32}\leftrightarrow
M_{\rm LEP}$ and $M_{10}$, are generated dynamically in a self-consistent
fashion.

The top-down scenario for string unification that we envision consists of the
following steps:

\begin{description}
\item (i) $Q=M_{\rm string}$: The SU(5), U(1), SO(10), and SU(4) gauge
couplings are unified at the common value $g$. String threshold corrections,
that may shift $M_{\rm string}$, are expected to be very small in this class of
models \cite{DF}.
\item (iia) $\Lambda_{10}<Q<M_{\rm string}$: The hidden SO(10) group evolves
according to the one-loop beta function $\beta_{10}=-24+N_{10}$, where $N_{10}$
is the number of SO(10) decaplets, and confines at the scale
$\Lambda_{10}=M_{\rm string}\, e^{8\pi^2/g^2\beta_{10}}\sim10^{15-16}\GeV$.
\item (iib) $\Lambda_4<Q<M_{\rm string}$: The hidden SU(4) group evolves
according to the one-loop beta function $\beta_4=-12+{1\over2}N_4+N_6$, where
$N_4$ is the number of \r{4} and \rb{4} fields, and $N_6$ is the number of
\r{6} fields, and confines at the scale
\begin{equation}
\Lambda_4=M_{\rm string}\, e^{8\pi^2/g^2\beta_4}\ .
\label{eq:Lambda4Formula}
\end{equation}
This scale depends on the detailed spectrum of \r{4},\rb{4},\r{6} particles
and on $g$. For concreteness, we will assume the typical case of $N_6=0$ and
$N_4=0,2,4$.
\item (iii) $M_{32}<Q<M_{\rm string}$: The SU(5) and U(1) gauge groups
evolve according to the following one-loop beta functions \cite{faspects}
\begin{eqnarray}
b_5&=&-15+2N_g+\coeff{1}{2}N_5+\coeff{3}{2}N_{10}=-2\ ,
\label{eq:b5}\\
b_1&=&2N_g+\coeff{1}{2}N_5+\coeff{1}{4}N_{10}+\coeff{5}{8}N_4=8+\coeff{5}{8}N_4
\ ,\label{eq:b1}
\end{eqnarray}
where $N_g=3$ is the number of generations, $N_5=2$ is the number of Higgs
pentaplets ($h,\bar h$), $N_{10}=4$ is the number of Higgs decaplets (two
pairs of (\r{10},\rb{10})), and $N_4$ is the number of (light) hidden
\r{4},\rb{4} fields, as in item (iib). (The hidden fields are SU(5)
singlets and do not affect the running of SU(5).) Symmetry breaking down to
the Standard Model gauge group occurs at the scale $M_{32}$, as triggered by
the the vevs $\vev{\nu^c_H}=\vev{\nu^c_{\bar H}}$ of the neutral components of
the (\r{10},\rb{10}) Higgs representations.
\item (iv) $M_{10}<Q<M_{32}$: The SU(3), SU(2), and $\rm U(1)_Y$ gauge
couplings evolve according to the one-loop beta functions \cite{faspects}
\begin{equation}
\left(\begin{array}{c} b_Y\\ b_2\\ b_3\end{array}\right)=
-\left(\begin{array}{c}0\\ 6\\ 9\end{array}\right)
+N_g\left(\begin{array}{c}2\\ 2\\ 2\end{array}\right)
+N_2\left(\begin{array}{c}{3\over10}\\ {1\over2}\\ 0\end{array}\right)
+N_3\left(\begin{array}{c}{1\over5}\\ 0\\ {1\over2}\end{array}\right)
+N_{32}\left(\begin{array}{c}{1\over10}\\ {3\over2}\\ 1\end{array}\right)
+N_4\left(\begin{array}{c}{3\over5}\\ 0\\ 0\end{array}\right)\ ,
\label{eq:bYb2b3}
\end{equation}
where $N_g=3$, $N_2=2$ is the number of light Higgs doublets, $N_3=2$ is the
number of $D^c,\bar D^c $ fields from the extra (\r{10},\rb{10}), and
$N_{32}=2$ the corresponding number of $Q,\bar Q$. Here $N_4$ is the number of
\r{4},\rb{4} hidden fields (as in item (iib)), which decouple
from the evolution at the scale $\Lambda_4$ (higher than $M_{10}$).
We obtain: $b_Y={36\over5}+{3\over5}N_4$, $b_2=4$, and $b_3=0$.\footnote{In
our numerical calculations below we include the full two-loop beta function
coefficients \cite{faspects}, which smooth out this zero-slope behavior.}
Because of the non-standard embedding of the electric charge generator in
SU(5), the gauge couplings at $M_{32}$ are related via $25/\alpha_Y=
1/\alpha_5+24/\alpha_1$ \cite{faspects}. We first adjust $M_{10}$ to obtain
string unification at $M_{\rm string}$, given values of the low-energy gauge
couplings and $N_4$, and later check the consistency of this calculation
against the dynamically determined value of $M_{10}$ (given in terms of
$\Lambda_4$, as discussed below).
\item (v) $M_Z<Q<M_{10}$: The extra (\r{10},\rb{10}) representations
decouple at $M_{10}$ and the SU(3), SU(2), and $\rm U(1)_Y$ gauge couplings
evolve according to the one-loop beta functions in Eq.~(\ref{eq:bYb2b3}). With
the traditional values $N_g=3$, $N_2=2$, and $N_3=N_{32}=0$, we obtain the
usual result $b_Y={33\over5}$, $b_2=1$, and $b_3=-3$.
\end{description}

The above general scenario requires the dynamical generation of three mass
scales: $M_{32}$, $M_{10}$, $\Lambda_4$. The generation of the latter scale is
well known, and was discussed in item (iib) above. Generation of the
scale of unified symmetry breaking $M_{32}$ has remained a puzzle in
string model-building. Here we advocate a mechanism of pure stringy origin.
Many (if not all) realistic string models possess a U(1) factor in the gauge
group whose trace over the massless string states does not vanish, \ie, an
``anomalous" $\rm U_A(1)$. It has been long known \cite{DSW} that this anomaly
is simply the result of truncating the low-energy effective theory to the
massless spectrum, and it is not present in the full string theory. Its
presence affects the D-term contribution to the scalar potential from this U(1)
gauge symmetry, and breaks supersymmetry in the ``original" vacuum.
Supersymmetry can be easily restored by sliding to a nearby vacuum, which is
still a consistent solution of string theory, and which is parametrized by
vevs of scalar fields of typical magnitude $M_{\rm string}$ or lower. In
particular, the set of shifted scalar fields includes those that break the
SU(5)$\times$U(1) gauge symmetry, which then become dynamically determined.
In specific models one finds that $M_{32}\approx g\vev{\nu^c_H}\sim M_{\rm
LEP}$ can be readily obtained \cite{Erice95}.

Our scenario also requires the dynamical generation of the  $M_{10}$ scale,
which provides masses to the extra (\r{10},\rb{10}) representations. We propose
to obtain this scale through a non-renormalizable superpotential coupling
of the form
\begin{equation}
\lambda\, (10)\,(\overline{10})\,(4)\,(\bar4)\,{1\over M}\ ,
\label{eq:NRT}
\end{equation}
as generically available in this class of string models, where
$M\approx10^{18}\GeV$ is the appropriate scale \cite{decisive}, and
$\lambda\lsim1$ is expected. Assuming that in the SU(4) condensation process
the $N_4$ hidden fields obtain masses ${\cal O}(\Lambda_4)$, the $\vev{4\bar4}$
condensate is estimated to be $\vev{4\bar4}\sim\Lambda^2_4$ and thus
\begin{equation}
M_{10}\sim \lambda\,{\vev{4\bar4}\over M}\sim {\Lambda^2_4\over M}\ .
\label{eq:M10}
\end{equation}

Whether the above scenario for string unification is realistic or not can be
determined by following the evolution of the gauge couplings from the bottom
up: starting from the well measured Standard Model gauge couplings and running
up to the string scale, using the two-loop renormalization group equations.
In practice, $M_{\rm 32}$ is determined by the low-energy gauge couplings
to be close to $M_{\rm LEP}$, irrespective of the value of $M_{10}$, whereas
$M_{\rm string}$ depends on $M_{10}$. We first adjust $M_{10}$ to obtain string
unification, and then check the consistency of our procedure against our
dynamical prediction for $M_{10}$ in Eq.~(\ref{eq:M10}).

\begin{figure}[p]
\vspace{6.5in}
\includegraphics{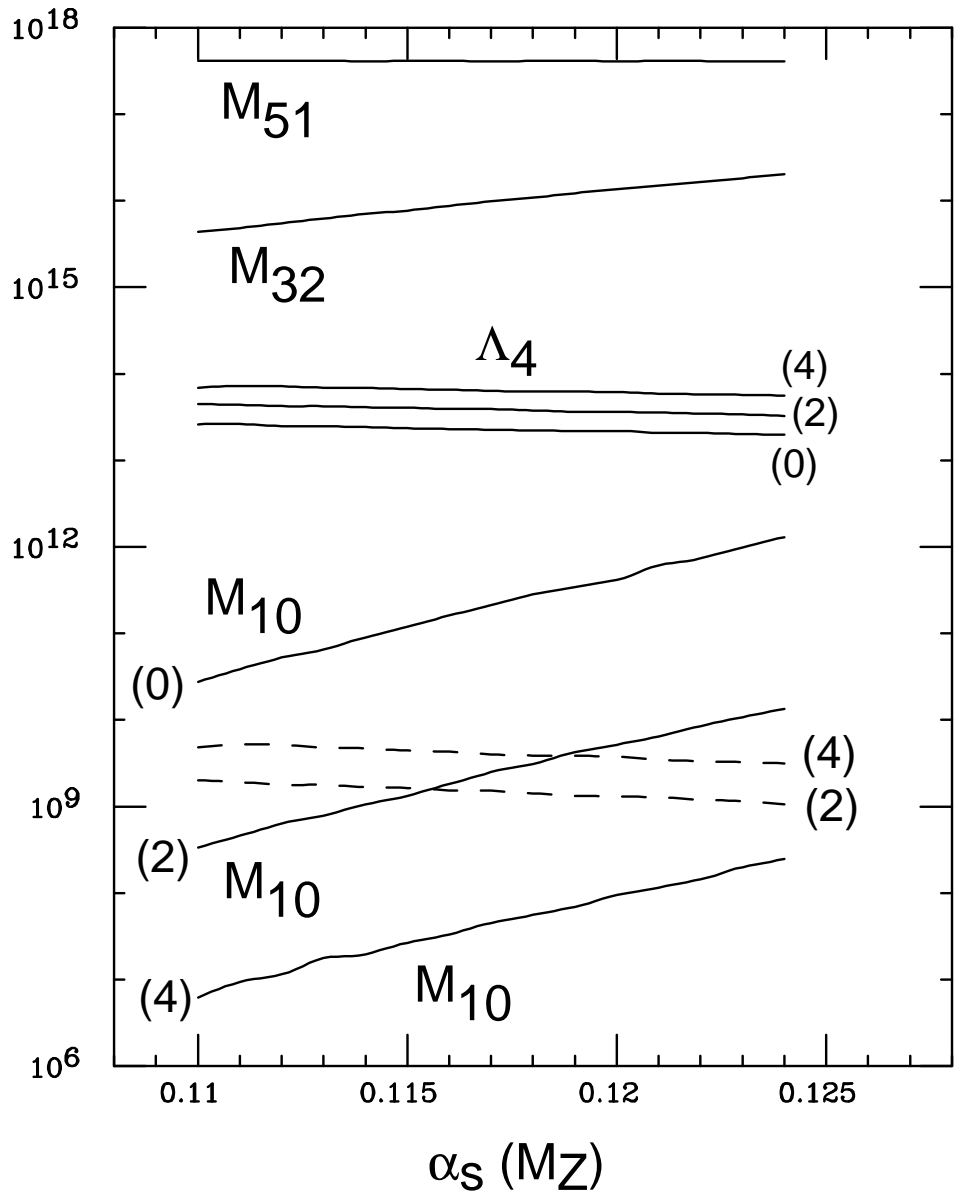}
\caption{The calculated values of the SU(5) and U(1) unification scale
($M_{51}$, identified with $M_{\rm string}$), the SU(5) unification scale
($M_{32}$, identified with $M_{\rm LEP}$), the SU(4) confinement scale
$\Lambda_4$, and the intermediate scale $M_{10}$, as a function of
$\alpha_s(M_Z)$ for $N_4=0,2,4$ (indicated in parenthesis). Dashed lines
display estimates of the dynamical prediction for $M_{10}$. Note that
$\alpha_s=0.116$ and $N_4=2$ work rather well. (All masses in GeV.)}
\label{fig:Scales}
\end{figure}

\begin{figure}[p]
\vspace{6.5in}
\includegraphics{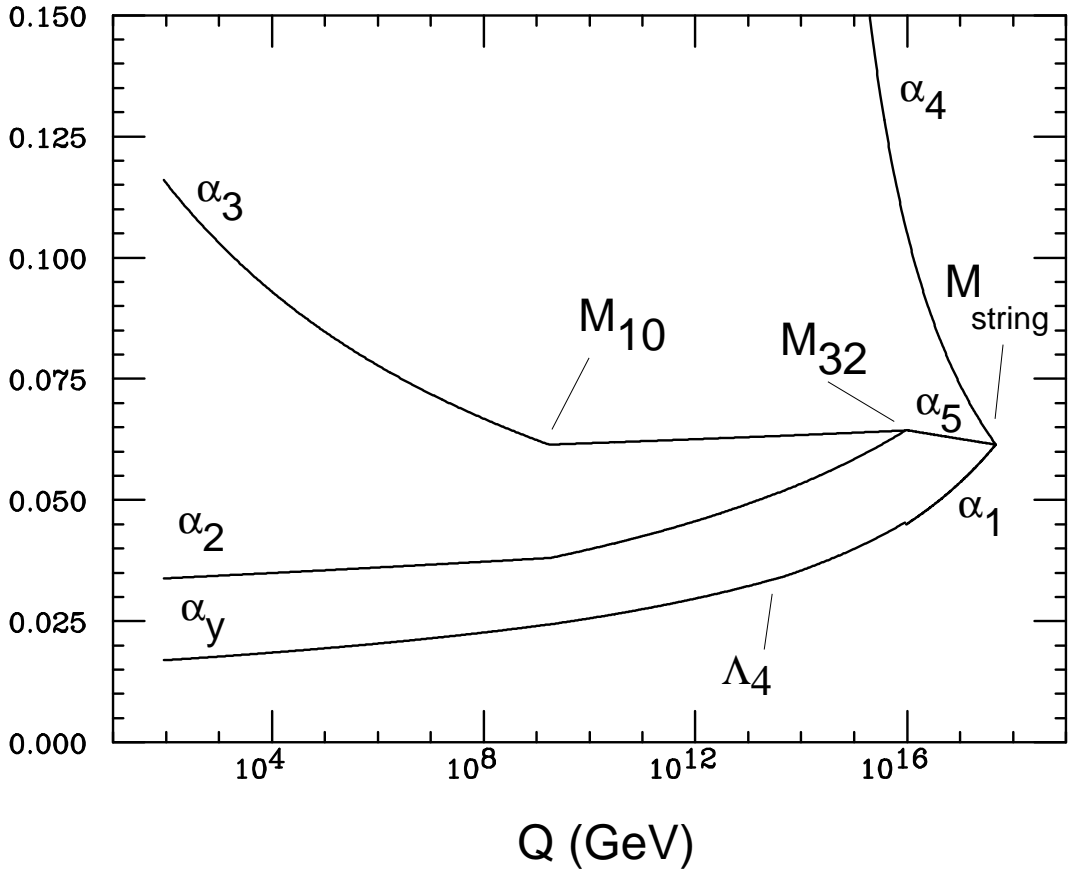}
\caption{The running of the gauge couplings for the preferred values
$\alpha_s(M_Z)=0.116$ and $N_4=2$. One obtains $M_{10}=1.8\times10^9$~GeV,
$M_{32}=8.7\times10^{15}$~GeV, $M_{51}=4.4\times10^{17}$~GeV,
$\Lambda_4=3.9\times10^{13}$~GeV, and $g=0.88$. This value of $M_{10}$ agrees
rather well with the dynamical prediction $M_{10}\sim10^9$~GeV.}
\label{fig:Runnings}
\end{figure}

In Fig.~\ref{fig:Scales} we show the calculated values of
$M_{51}\leftrightarrow M_{\rm string}$, $M_{\rm 32}$, $\Lambda_4$,
and $M_{10}$ as a function of $\alpha_s(M_Z)$, calculated to two-loop
precision.\footnote{Note that $\Lambda_4$ does not decrease with increasing
$N_4$ (as naively expected) because of its dependence on $g$, which needs
to be self-consistently determined (in the bottom-up approach) for every choice
of $N_4$, and which increases with $N_4$. This is also the source of the
$\alpha_s$ dependence of $\Lambda_4$.} The realistic case requires $N_4\not=0$
to be able to generate $M_{10}$ dynamically. Different values of $N_4$ affect
$M_{10}$, as shown in the figure for $N_4=0,2,4$. Comparing these
calculated values of $M_{10}$ with those predicted from Eq.~(\ref{eq:M10})
(dashed lines in Fig.~\ref{fig:Scales}), shows that $N_4=2$ for
$\alpha_s=0.116$ works rather well: $M_{10}\sim 10^9\GeV$. It is most
interesting that such self-consistency checks work at all, and that they
can constrain the spectrum of hidden sector states with observable-sector
quantum numbers. We finally put all pieces together and show in
Fig.~\ref{fig:Runnings} the running of the gauge couplings for the favored
values $\alpha_s=0.116$ and $N_4=2$, with all the scales as indicated.

Some phenomenological aspects of this type of scenario (restricted to
$Q<M_{32}$), including the (small) effects of light and GUT thresholds and a
possibly observable proton decay signal into $e^+\pi^0$ at Superkamiokande,
have been recently explored in Ref.~\cite{lowering}. The above scenario should
motivate further flipped SU(5) string model-building along these lines.

\newpage

\section*{Acknowledgments}
The work of J.~L. has been supported in part by DOE grant DE-FG05-93-ER-40717.
The work of D.V.N. has been supported in part by DOE grant DE-FG05-91-ER-40633.

\clearpage

\clearpage

\end{document}